\documentclass{ifacconf}

\usepackage{graphicx}      
\usepackage{natbib}        

\usepackage{amsmath}
\usepackage{amsfonts}
\usepackage{eurosym}
\usepackage{xcolor}
\usepackage{siunitx}
\usepackage{float}
\usepackage{tabularx, booktabs}

\begin{document}
\begin{frontmatter}

\title{RL-Guided MPC for \\Autonomous Greenhouse Control} 

\author[First]{Salim Msaad} 
\author[First]{Murray Harraway} 
\author[First]{Robert D. McAllister}

\address[First]{Delft Center for Systems and Control (DCSC),\\
Delft University of Technology, Delft, The Netherlands}


\begin{abstract} 
The efficient operation of greenhouses is essential for enhancing crop yield while minimizing energy costs. 
This paper investigates a control strategy that integrates Reinforcement Learning (RL) and Model Predictive Control (MPC) to optimize economic benefits in autonomous greenhouses. 
Previous research has explored the use of RL and MPC for greenhouse control individually, or by using MPC as the function approximator for the RL agent.
This study introduces the RL-Guided MPC framework, where a RL policy is trained and then used to construct a terminal cost and terminal region constraint for the MPC optimization problem.
This approach leverages the ability to handle uncertainties of RL with MPC's online optimization to improve overall control performance.
The RL-Guided MPC framework is compared with both MPC and RL via numerical simulations.
Two scenarios are considered: a deterministic environment and an uncertain environment.
Simulation results demonstrate that, in both environments, RL-Guided MPC outperforms both RL and MPC with shorter prediction horizons.
\end{abstract}


\end{frontmatter}


\section{Introduction}\label{sec:introduction}
The sustainable and efficient operation of greenhouses is pivotal to modern agriculture, offering a controlled environment to maximize crop yield while optimizing resource efficiency. 
Modern greenhouse systems demand advanced control strategies to dynamically manage climatic variables, such as temperature, humidity, and CO2 levels, with fluctuating external conditions. 
Traditional control methods struggle to balance long-term economic objectives with real-time operational demands. 
Reinforcement Learning (RL) and Model Predictive Control (MPC) have shown promise in addressing these challenges.
Prior research has explored RL and MPC as standalone solutions for greenhouse automation. 
RL excels in handling stochastic environments, while MPC leverages model-based optimization to enforce constraints and recalibrate actions in real time. 
\\
RL approaches for greenhouse control have been studied in \cite{vanlaatum_2024} and in \cite{morcego_reinforcement_2023}.
The first study presents an open-source RL environment for greenhouse control, comparing different RL approaches, while the second study introduces a RL-based controller that utilizes deep deterministic policy gradient for greenhouse climate control, comparing it to a MPC controller.
\cite{boersma_robust_2022} propose a robust MPC controller for greenhouse climate control, showing improved performance over traditional MPC. 
\cite{mallick_reinforcement_2025} integrate RL and MPC by using MPC as a function approximator within the RL framework.
\\
The autonomous greenhouse control problem is an economic optimization task. 
Conventional MPC approaches focus on tracking an economically optimal steady-state for the system. 
However, for applications such as greenhouse control, steady-state tracking is not desirable as optimal economic operation requires dynamically adjusting the temperature, humidity, and CO2 levels in the greenhouse.  
Economic MPC, which uses economic costs directly in the stage cost and does not focus on stabilizing a steady-state target, is therefore more appropriate for greenhouse control. 
Nonetheless, a steady-state or reference trajectory is often used to improve and guarantee performance of economic MPC implementations \citep{amrit_economic_2011, risbeck_economic_2020, angeli_average_2012}
\\
This paper presents the RL-Guided MPC framework, where an RL policy informs the terminal cost and region constraints in the MPC optimization. 
The policy is used to build a cost function approximator for the terminal cost, while rollouts using the actor define the terminal region constraint.
This integration improves control performance while ensuring computational efficiency in scenarios with limited prediction horizons.
The efficacy of the proposed approach is validated through numerical simulations, comparing RL-Guided MPC against standalone MPC and RL controllers. 
The simulations are conducted in two environments.
One environment is deterministic, while the other is stochastic with parametric uncertainty.
\\
Section~\ref{sec:background} outlines the greenhouse model, simulation environments, and the optimization problem tackled 
by the RL agent, MPC, and RL-Guided MPC. 
Section~\ref{sec:reinforcement_learning} and Section~\ref{sec:model_predictive_control} describe the RL agent and MPC 
controller, while Section~\ref{sec:RL-guided_MPC} introduces the RL-Guided MPC framework. 
Numerical results are presented in Section~\ref{sec:results}, and the conclusions are presented in 
Section~\ref{sec:conclusions}.


\begin{table*}[h]
	\caption{Physical meaning of input $u$, state $x$, and disturbance $d$}\label{tab:physical_meaning_variables}
	\centering
	\begin{tabular}{ll|ll|ll}
		\hline
		\rule{0pt}{10pt}
		$u_1$ & CO$_2$ injection (\si{\milli\gram\per\meter\squared\per\second}) & $x_1$ & dry weight (\si{\kilo\gram\per\meter\squared}) & $d_1$ & solar radiation (\si{\watt\per\meter\squared}) \\
		\rule{0pt}{0pt}
		$u_2$ & ventilation (\si{\milli\meter\per\second}) & $x_2$ & indoor CO$_2$ (\text{ppm}) & $d_2$ & outdoor CO$_2$ (\si{\kilo\gram\per\meter\cubed}) \\
		\rule{0pt}{0pt}
		$u_3$ & heating (\si{\watt\per\meter\squared}) & $x_3$ & indoor temperature (\si{\celsius}) & $d_3$ & outdoor temperature (\si{\celsius}) \\
		\rule{0pt}{0pt}
		 &  & $x_4$ & indoor humidity (\%) & $d_4$ & outdoor humidity (\si{\kilo\gram\per\meter\cubed}) \\[3pt]
		\hline
	\end{tabular}
\end{table*}
\section{Background}\label{sec:background}

\subsection{Greenhouse Model}\label{sec:greenhouse_model}
The greenhouse and crop model utilized in this paper is the same as in \cite{van_henten_greenhouse_1994}.
The model is discretised with the fourth order Runge-Kutta method with a sample period \(\Delta t = 1800\) s (30 minutes), resulting in the following state space model:
\begin{equation}\label{eq:greenhouse_model_discrete}
	x(k+1) = f(x(k),u(k),d(k),p)
\end{equation}
with discrete time \(k \in \mathbb{Z}^{+}\), state variable \(x(k) \in \mathbb{R}^4\), control input \(u(k) \in \mathbb{R}^3\) and weather disturbance \(d(k) \in \mathbb{R}^4\).
The parameter \(p \in \mathbb{R}^{22}\) represents all parameters used in the model.
The values and meaning of \(p\) and the nonlinear function \(f(\cdot)\) are described in \cite{boersma_robust_2022}.
State estimation is not considered in this work.
It is assumed that the state variable \(x(k)\) is always known.
Table~\ref{tab:physical_meaning_variables} provides the physical meaning of the state, input, and disturbance variables.
\\
The weather data for simulations and training was sourced from the Venlow Greenhouse in Bleiswijk, covering the period from January 30 to March 11, 2014.
The weather is assumed to be deterministic and known at each time step. 
This dataset spans a 40-day growing period with a time step of 30 minutes, resulting in a total of 1920 time steps.
For the numerical simulations, two scenarios are considered: a deterministic environment and a stochastic environment.
In a real-world greenhouse, uncertainty is present in all aspects of the model.
Uncertainty may arise in the weather prediction, measurement noise on the outputs, and the control inputs may not be exact.
As done in \cite{boersma_robust_2022}, the uncertainty in the stochastic environment is modelled as parametric uncertainty, which aims to offer a simplified representation of the real-world uncertainties.
It is assumed that the uncertain parameters \(\hat p \in \mathbb{R}^{22}\) follow the uniform probability distribution
\begin{equation}\label{eq:uncertain_parameters}
	\hat p \sim \mathcal{U}( p(1-\delta), p(1+\delta) ),
\end{equation}
where vector \(p\) represents the nominal values of the parameters and $\delta$ is a percentage that defines the range of the distribution.
The value of $\delta$ is set to 5\% for all of the stochastic simulations presented in Section~\ref{sec:results}.
For each simulation in the stochastic environment, a new set of uncertain parameters \(\hat p\) is sampled from the distribution in (\ref{eq:uncertain_parameters}), and the system dynamics in (\ref{eq:greenhouse_model_discrete}), with $p=\hat p$, are used for simulation.

\begin{table}
	\caption{Pricing and penalty factors}\label{tab:pricing_and_penalty_factors}
	\centering
	\begin{tabular}{lll}
		\hline
		\rule{0pt}{8pt}
		Symbol & Value & Unit \\
		\hline
		\rule{0pt}{8pt}
		$c_1$ & $1.906 \cdot 10^{-1}$ & \euro\,\si{\per\milli\gram} \\
		\rule{0pt}{0pt}
		$c_3$ & $1.281 \cdot 10^{-1}$ & \euro\,\si{\per\joule} \\
		\rule{0pt}{0pt}
		$r$ & $20.93$\textsuperscript{*} & \euro\,\si{\per\kilo\gram} \\[2pt]
		\hline
		\rule{0pt}{8pt}
		$\lambda_2$ & $5 \cdot 10^{-5}$ & \euro (ppm \si{\meter\squared})$^{-1}$ \\
		\rule{0pt}{0pt}
		$\lambda_3$ & $5 \cdot 10^{-3}$ & \euro (\si{\per\degree C \,\meter\squared})$^{-1}$ \\
		\rule{0pt}{0pt}
		$\lambda_4$ & $7 \cdot 10^{-4}$ & \euro $ (\% \, \text{m}^2)^{-1}$ \\[2pt]
		\hline
	\end{tabular}

	\vspace{4pt}
	\raggedright\textsuperscript{*} Corresponding to 1.07 \euro\,\si{\per\kilo\gram} for fresh weight.
\end{table}

\begin{table}
	\caption{Input and output constraints and initial conditions}\label{tab:inputs_and_outputs_constraints}
	\centering
	\begin{tabular}{ll|ll|ll}
		\hline
		\rule{0pt}{8pt}
		Symbol & Value & Symbol & Value & Symbol & Value \\
		\hline
		\rule{0pt}{8pt}
		$u^\text{min}_1$ & 0 & $u^\text{max}_1$ & 1.2 & $u_1(0)$ & 0 \\
		\rule{0pt}{0pt}
		$u^\text{min}_2$ & 0 & $u^\text{max}_2$ & 7.5 & $u_2(0)$ & 0 \\
		\rule{0pt}{0pt}
		$u^\text{min}_3$ & 0 & $u^\text{max}_3$ & 150 & $u_3(0)$ & 50 \\[2pt]
		\hline
		\rule{0pt}{8pt}
		$x^\text{min}_1$ & 0 & $x^\text{max}_1$ & $\infty$ & $x_1(0)$ & 0.0035 \\
		\rule{0pt}{0pt}
		$x^\text{min}_2$ & 500 & $x^\text{max}_2$ & 1600 & $x_2(0)$ & 0.001 \\
		\rule{0pt}{0pt}
		$x^\text{min}_3$ & 10 & $x^\text{max}_3$ & 20 & $x_3(0)$ & 15 \\
		\rule{0pt}{0pt}
		$x^\text{min}_4$ & 0 & $x^\text{max}_4$ & 80\% (78\%)\textsuperscript{*} & $x_4(0)$ & 0.008 \\[2pt]
		\hline
	\end{tabular}

	\vspace{4pt}
	\raggedright\textsuperscript{*} Constraint tightening of 2\% for the stochastic case.
\end{table}

\subsection{Optimization Problem}\label{sec:optimization_problem}
Optimal greenhouse control involves maintaining suitable environmental conditions for crop growth, including temperature, humidity, and CO$_2$ levels, while minimizing resource use.
This is achieved through control inputs for heating using heating pipes, ventilation by controlling window openings, and CO$_2$ injection.
The goal is to maximize crop yield while minimizing resource consumption. 
Deciding the harvest time is also an important factor in profitability. 
However, in this study, a fixed growing period of 40 days is chosen instead of including this decision in the optimization problem. 
This approach aligns with other studies in the literature (\cite{boersma_robust_2022, morcego_reinforcement_2023, mallick_reinforcement_2025}).
At the conclusion of this growing period, the crop is harvested and sold, marking the end of the cultivation cyle.
The revenue generated from the sale is then used to calculate the Economic Profit Indicator ($EPI$), which is determined by subtracting the heating and CO$_2$ costs from the total earnings, as follows:
\begin{equation}\label{eq:EPI}
	EPI = r\, x_1(t_f) - \sum_{k=0}^{t_f} \bigl( c_1 u_1(k) + c_3 u_3(k) \bigr) \Delta t,
\end{equation}
where $t_f$ denotes the final time, set at 40 days, $c_1$ represent the CO$_2$ injection cost coefficient, $c_3$ the heating cost coefficient, $r$ the revenue coefficient from lettuce sales, and $\Delta t$ the sample period.
The values of these pricing factors are listed in Table~\ref{tab:pricing_and_penalty_factors}.
\\
The $EPI$ defines the objective function for the greenhouse control problem.
However, directly optimizing (\ref{eq:EPI}) using MPC or RL is difficult due to the sparse reward structure, since rewards are given only at the end of the growing period.
To overcome this challenge, the following economic stage cost is defined for each time step:
\begin{equation}\label{eq:stage_cost}
	\begin{aligned}
		\ell_e\bigl( x(k),u(k) \bigr) = 
		& \, - r \bigl( x_1(k) - x_1(k-1) \bigr) \\
		& \, + \bigl( c_1 u_1(k) + c_3 u_3(k) \bigr) \Delta t.
	\end{aligned}
\end{equation}
The cumulative sum of the stage costs in (\ref{eq:stage_cost}) over the growing period is equivalent to the $EPI$ in (\ref{eq:EPI}), but with the opposite sign.
State constraints are essential to ensure the system operates within realistic and feasible bounds. 
Temperature, humidity, and CO$_2$ concentration must stay within specified minimum and maximum limits. 
However, due to inherent uncertainties, these constraints cannot be strictly enforced. 
Instead, they are treated as soft constraints by incorporating three penalty terms into the objective function to account for any violations.
As a result, the following stage cost function is defined:
\begin{equation}\label{eq:objective_function}
	\begin{aligned}
		\ell\bigl(x(k), u(k)\bigr) = \, &\ell_e\bigl( x(k), u(k) \bigr) + g_2\bigl(x_2(k)\bigr)\\
		& + g_3\bigl(x_3(k)\bigr) + g_4\bigl(x_4(k)\bigr),
	\end{aligned}
\end{equation}
where $g_2$, $g_3$, $g_4$ are the penalty functions for the indoor CO$_2$ concentration, temperature, and humidity, respectively.
These penalty functions are defined as
\begin{equation}\label{eq:penalty_factors}
	g_i\bigl(x_i(k)\bigr) = 
	\begin{cases}
		\lambda_i \bigl(x_i(k) - x_i^\text{max}\bigr) \ &\text{if} \ x_i(k) > x_i^\text{max}, \\
		\lambda_i \bigl(x_i^\text{min} - x_i(k)\bigr) \ &\text{if} \ x_i(k) < x_i^\text{min}, \\
		0 &\text{otherwise}.
	\end{cases}
\end{equation}
The penalty coefficients $\lambda_2$, $\lambda_3$, and $\lambda_4$ are defined in Table~\ref{tab:pricing_and_penalty_factors}. 
The state's minimum and maximum threshold values $x_{i}^\text{min}$ and $x_{i}^\text{max}$ are defined in Table~\ref{tab:inputs_and_outputs_constraints}.
In addition to the state constraints, the control inputs are also subject to constraints.
The heating injection, ventilation, and CO$_2$ injection are bounded by their respective minimum and maximum values, as defined in Table~\ref{tab:inputs_and_outputs_constraints}.
Moreover, the control inputs are subject to rate constraints to prevent abrupt changes in the control actions.
For each control input $i=\{1,2,3\}$, the absolute rate of change is limited to $\delta u_i^\text{max} = u_i^\text{max}/10$.
\\
Considering the objective function in (\ref{eq:objective_function}), along with the state and rate constraints, the optimal greenhouse control task is described by the following optimization problem: 
\begin{subequations}\label{eq:optimization_problem}
	\begin{align}
		& \underset{u(\cdot)}{\text{min}} 
		& & \sum_{k=0}^{t_f} \ell\bigl( x(k),u(k) \bigr), \label{eq:optimization_problem:objective_function} \\
		& \, \text{s.t.}
		& & x(k+1) = f\bigl(x(k),u(k),d(k),p\bigr), \label{eq:optimization_problem:state_function}\\
		& & & u_\text{min} \leq u_i(k) \leq u_\text{max}, \label{eq:optimization_problem:input_constraint}\\
		& & & \mspace{-4mu} \left| u(k) - u(k-1) \right| \leq \delta u_\text{max}. \label{eq:optimization_problem:input_rate_constraint}
	\end{align}
\end{subequations}

The RL agent, MPC and RL-Guided MPC all seek to solve the optimization problem in (\ref{eq:optimization_problem}) and are ultimately evaluated based on the cumulative cost in (\ref{eq:optimization_problem:objective_function}).
The initial conditions, detailed in Table~\ref{tab:inputs_and_outputs_constraints}, were kept constant for every episode and for both the deterministic and stochastic cases.


\section{Reinforcement Learning}\label{sec:reinforcement_learning}
Actor-critic methods combine elements of both policy-based and value-based approaches, enabling stable and efficient learning. 
The actor is responsible for learning a policy, while the critic evaluates the quality of actions taken. 
By leveraging this dual structure, actor-critic algorithms allow policies to improve based on both direct experience and feedback from the critic.
One key advantage of actor-critic algorithms is their ability to handle continuous state and action spaces effectively (\cite{sutton_reinforcement_2020}),
in contrast to other methods, such as Q-learning, that are better suited to discrete environments. 
This makes them well-suited for complex control tasks, such as greenhouse climate optimization.
The actor-critic algorithm employed in this study is soft actor-critic (SAC) (\cite{haarnoja_SoftActorCritic_2018}), a widely used actor-critic algorithm.
SAC introduces an entropy term into the reward function, explicitly balancing exploration and exploitation. 
This leads to the learning of a policy $\pi_\theta(s(k))$, function of the current observation $s(k)$ and parameterized by $\theta$, where the entropy term determines policy randomness. 
By optimizing both the expected cumulative reward and entropy, the agent encourages diverse action selection.
\\
Careful selection of the agent's observation space is crucial for effective learning. 
Providing insufficient information may hinder the agent's ability to learn meaningful strategies, while excessive information can make it difficult to extract relevant patterns.
In a real-world greenhouse setting, expert growers do not have direct access to the current dry weight of the lettuce crop. 
However, in this study, it is assumed that this value, denoted by $x_1(k)$, is available to the RL agent, to MPC and to RL-Guided MPC. 
The agent also receives data on the greenhouse's indoor temperature $x_2(k)$, CO$_2$ concentration $x_3(k)$, and humidity $x_4(k)$, which are typically measured by sensors in real-world greenhouses.
Moreover, the previous input $u(k-1)$ is included in the observation space.
This information is crucial, since the difference between the current and previous control inputs is constrained, by (\ref{eq:optimization_problem:input_rate_constraint}), to prevent abrupt changes.
The weather disturbance $d(k)$ is also available to the agent, as it is for the MPC and for the RL-Guided MPC.
Finally, the agent is designed to be time-aware, meaning that the current time step $k$ is explicitly provided. 
This allows the agent to make decisions based on the stage of the growing period.
Without this information, the agent would not be able to distinguish between early and late stages of the growing period, which is crucial for effective control.
\begin{table}
	\caption{Hyperparameters for the RL agent}\label{tab:RL-hyperparameters}
	\centering
	\begin{tabular}{ll}
		\hline
		\rule{0pt}{8pt}
		Parameter & Value \\
		\hline
		\rule{0pt}{8pt}
		Training episodes & 100 \\
		\rule{0pt}{0pt}
		Warm-up episodes & 9 \\
		\rule{0pt}{0pt}
		Hidden layers & 2 \\
		\rule{0pt}{0pt}
		Neurons per hidden layer & 128 \\
		\rule{0pt}{0pt}
		Batch size & 1024 \\
		\rule{0pt}{0pt}
		Learning rate & $5 \cdot 10^{-3}$ \\
		\rule{0pt}{0pt}
		Buffer size & 100,000 \\
		\rule{0pt}{0pt}
		Activation function & \texttt{ReLU} \\
		\rule{0pt}{0pt}
		Discount factor $\gamma$ & 0.95 \\
		\hline
	\end{tabular}
\end{table}
Therefore, the observation space is defined as:
\begin{equation}\label{eq:observation_space} 
	s(k) = \bigl(x(k), u(k-1), d(k), k\bigr).
\end{equation}
To ensure that the control input $u(k)$ adheres to the constraints specified in (\ref{eq:optimization_problem}), the agent's action, denoted as $a(k) = \pi_\theta(s(k))$ with $a \in [-1,1]^3$, is interpreted as an adjustment to the previous control input:  
\begin{equation}\label{eq:action_to_input}
	\begin{aligned}
		u(k) = \, \text{max} \bigl( & u_\text{min}, \\
		& \min (u_\text{max}, u(k-1) + a(k) \delta u_\text{max}) \bigr).
	\end{aligned}
\end{equation}  
The reward function that the RL agent is trained to maximize is the same objective function defined in (\ref{eq:objective_function}) but with an inverted sign.
This formulation ensures that both methods operate under a consistent optimization framework, allowing for direct comparisons between all approaches.
The discount factor, $\gamma$, is a crucial hyperparameter. 
For long-term tasks, $\gamma = 1$ allows the agent to consider the entire growth period and optimize long-term economic outcomes. 
However, it can destabilize training. 
To address this, we used $\gamma = 0.95$, which improved policy effectiveness. 
Table~\ref{tab:RL-hyperparameters} lists all RL agent hyperparameters.
\\
Two different RL agents were trained, one for the deterministic environment and the other for the stochastic environment.
Both agents were trained using identical hyperparameters, with the only difference being the environment, and consequently, the datasets they were exposed to during training.


\section{Model Predictive Control}\label{sec:model_predictive_control}
The MPC controller in this study is designed to solve the optimization problem defined in (\ref{eq:optimization_problem}).
At time step $k_0$, the following optimization problem is solved:
\begin{subequations}\label{eq:MPC_optimization_problem}
	\begin{align}
		& \underset{u(k)}{\text{min}}
		& & \sum_{k=k_0}^{k_0+N_p} \ell\bigl( x(k),u(k) \bigr), \\
		& \mspace{2mu} \text{s.t.}
		& & x(k+1) = f(x(k),u(k),d(k),p), \\
		& & & u_\text{min} \leq u(k) \leq u_\text{max}, \\
		& & & \mspace{-4mu} \left|u(k) - u(k-1)\right| \leq \delta u_\text{max}, \\
		& & & x(k_0) = x_{k_0},
	\end{align}
\end{subequations}
where $N_p$ is the prediction horizon, and $x_{k_0}$ is the system's state at time $k_0$.
Once the optimization problem is solved, the first control input is applied to the system, and the optimization problem is solved again at the next time step.
This process is repeated at each time step until the end of the growing period.
Moreover, to help with the feasibility of the optimization problem, the solver is warm-started with the previous solution.
The optimization problem is solved numerically using the open-source software framework CasADi (\cite{casadi}) and the IPOPT solver (\cite{ipopt}).
\\
Two different MPC controllers are designed, one for the deterministic environment and the other for the stochastic environment.
The MPC formulation in (\ref{eq:MPC_optimization_problem}) is used for both environments. 
The only modification for the stochastic case is a tighter indoor humidity constraint \( x^\text{max}_4 \), 78\% instead of 80\%.
This constraint tightening in the MPC formulation reduces the risk of violating the actual constraint at 80\% in the stochastic simulation environment.


\section{RL-Guided MPC}\label{sec:RL-guided_MPC}
In the following, we outline the RL-Guided MPC framework, which incorporates the agent trained in Section~\ref{sec:reinforcement_learning} into the MPC optimization problem of Section~\ref{sec:model_predictive_control}. 
First, we outline how the terminal cost is derived from the critic. 
Next, we describe how the actor is utilized to define the terminal region constraint. 
Finally, we illustrate how the actor helps to provide a warm-start initial guess for the RL-Guided MPC optimization problem.

\subsection{Terminal cost}
In the RL-Guided MPC framework, the terminal cost is directly obtained from the actor network of the SAC agent trained in Section~\ref{sec:reinforcement_learning}.
Using the learned policy $\pi_\theta$, multiple closed-loop trajectories are generated.
This data is then used to train a cost function approximator $\tilde J_\phi$ that estimates the expected return for a given state.
This approach enables the RL-Guided MPC to account for long-term effects over the entire growing period without extending the prediction horizon.
\\
The cost function approximator is trained through expected return learning.
Given a nominal state trajectory $x^\text{n}(k)$, 1000 time steps are uniformly sampled from the interval $\{0, 1, \dots, t_f\}$. 
For each sampled time step $k^{(i)}$, an initial state $x(k^{(i)})$ is sampled from a uniform distribution within a range defined by the nominal trajectory:  
\begin{equation}
	x(k^{(i)}) \sim \mathcal{U}\bigl(x^\text{n}_\text{min}(k^{(i)}), x^\text{n}_\text{max}(k^{(i)}) \bigr)
\end{equation}
where  
\begin{equation}
	\begin{aligned}
		x^\text{n}_\text{min}(k) &= x^\text{n}(k) (1-\sigma), \\
		x^\text{n}_\text{max}(k) &= x^\text{n}(k) (1+\sigma).
	\end{aligned}
\end{equation}
The parameter \( \sigma \), set at $50\%$, determines the spread of the sampled initial states around the nominal trajectory.  
\\
Next, from each sampled state $x(k^{(i)})$, the respective agent's observation $s(k^{(i)})$, defined in (\ref{eq:observation_space}), is constructed and the policy $\pi_\theta$ is used to obtain a closed-loop trajectory that extends until the end of the growing cycle $t_f$.
The cumulative cost of each trajectory is then defined as:
\begin{equation}
	J_{\pi_\theta}(s(k^{(i)})) = \sum_{k=k_s}^{t_f} \ell\bigl( x(k), \pi_\theta(s(k)) \bigr).
\end{equation}
The cumulative cost of each trajectory is collected together with its respective dry weight $x_1(k^{(i)})$ and sampled time $k^{(i)}$ to form dataset $\mathcal{D}$.
This dataset is divided into training and validation sets, with 80\% of the data used for training and 20\% for validation.
To construct the cost function approximator $\tilde J_{\phi}$, a neural network parameterized by $\phi$ is trained using the values of $x_1(k^{(i)})$ and $k^{(i)}$ as inputs and the values of $J_{\pi_\theta}(s(k^{(i)}))$ as targets.
The following loss function is minimized using the Adam optimizer (\cite{adam_optimizer_2017}):
\begin{equation*}
	\mathcal{L}(\phi, \mathcal{D}) = 
		\frac{1}{N_s} \sum_{i=1}^{N_s} \left( \tilde J_\phi(x_1(k^{(i)}), k^{(i)}) - J_{\pi_\theta}(s(k^{(i)})) \right)^2.
\end{equation*}
The hyperparameters used for the cost function approximator are indicated in Table~\ref{tab:NN-hyperparameters}. 
\begin{table}
	\caption{Hyperparameters for the cost function approximator}\label{tab:NN-hyperparameters}
	\centering
	\begin{tabular}{ll}
		\hline
		\rule{0pt}{8pt}
		Parameter & Value \\
		\hline
		\rule{0pt}{8pt}
		Hidden layers & 2 \\
		\rule{0pt}{0pt}
		Neurons per hidden layer & 128 \\
		\rule{0pt}{0pt}
		Batch size & 1024 \\
		\rule{0pt}{0pt}
		Learning rate & $1 \cdot 10^{-3}$ \\
		\rule{0pt}{0pt}
		Buffer size & 1024 \\
		\rule{0pt}{0pt}
		Activation function & \texttt{tanh} \\
		\hline
	\end{tabular}
\end{table}
\\
The cost function approximator is then added as a term to the objective function of the MPC optimization problem defined in (\ref{eq:MPC_optimization_problem}).
At each time step $k_0$, the cost function for the RL-Guided MPC is thus defined as
\begin{equation}
	\sum_{k=k_0}^{k_0+N_p} \ell\bigl( x(k),u(k) \bigr) + \tilde J_\phi\bigl( x_1(k_0+N_p), k_0+N_p \bigr).
\end{equation}

\subsection{Terminal region constraint and initial guess}
For the first time step, the terminal region constraint and the initial guess are constructed using the initial conditions, defined in Table~\ref{tab:inputs_and_outputs_constraints}, and the actor's policy.
The agent's observation $s(0)$ is constructed using the initial conditions.
Policy $\pi_\theta$ is then used to compute the closed-loop trajectory from the initial time step to $N_p+1$:
\begin{equation}\label{eq:first_observation_trajectory}
	\{ s(0), s(1), \dots, s(N_p+1) \},
\end{equation}
and its respective control input trajectory:
\begin{equation}\label{eq:first_input_trajectory}
	\{ u(0), u(1), \dots, u(N_p) \}.
\end{equation}
where $u(k) = \pi_\theta(s(k))$.
From the computed closed-loop trajectory in the observation space defined in (\ref{eq:first_observation_trajectory}), the respective state trajectory is obtained:
\begin{equation}
	\{ x(0), x(1), \dots, x(N_p+1) \}.
\end{equation}
The center of the terminal region constraint for the next optimization problem is then defined as the last state in this trajectory, i.e. $x_f = x(N_p+1)$.
The terminal region constraint for this initial optimization problem is then defined as
\begin{equation}\label{eq:terminal_region_constraint}
	[(1-\epsilon)x_f, (1+\epsilon)x_f],
\end{equation}
where $\epsilon$, expressed as a percentage, defines the size of the terminal region around $x_f$.
For the warm-start of the next optimization problem, the last $N_p$ elements of the input trajectory defined in (\ref{eq:first_input_trajectory}) are used as the initial guess.
\\
This terminal region constraint and initial guess construction is only used for the first time step.
For subsequent time steps, the terminal region constraint and initial guess are constructed using the solution of the previous optimization problem and the actor's policy.
The solution of the RL-Guided MPC optimization problem at each time step $k_0$ is denoted as
\begin{equation}
	\{u(k_0), u(k_0+1), \dots, u(k_0+N_p-1)\}.
\end{equation}
and the corresponding predicted states are denoted as
\begin{equation}
	\{x(k_0+1), x(k_0+2), \dots, x(k_0+N_p)\}.
\end{equation}
The terminal region constraint and the initial guess for the next optimization problem are constructed using these solutions and the actor's policy.
Starting from the final predicted state $x(k_0+N_p)$, the agent's observation $s(k_0+N_p)$ is constructed and policy $\pi_\theta$ is used to obtain the center of the terminal region for the next optimization problem:
\begin{equation}\label{eq:center_terminal_region}
	x_f = f(x(k_0+N_p), \pi_\theta(s(k_0+N_p)), d(k_0+N_p), p).
\end{equation}
The terminal region constraint for the next optimization problem is then defined as in (\ref{eq:terminal_region_constraint}).
The initial guess for the next optimization problem is constructed as
\begin{equation}
	\begin{aligned}
		\bigl\{ u(k_0+1), &u(k_0+2), \dots, \\
		& u(k_0+N_p-1), \pi_\theta \bigl( s(k_0+N_p) \bigr) \bigr\}.
	\end{aligned}
\end{equation}
Two RL-Guided MPC controllers were developed, one for the deterministic environment and the other for the stochastic environment.
These controllers differ in the policy used to construct the terminal region constraint and the initial guess.
Moreover, for the stochastic case, the indoor humidity constraint \( x^\text{max}_4 \) is tightened to 78\% to account for model uncertainties, reducing the risk of constraint violations, as for the MPC for the stochastic case in Section~\ref{sec:model_predictive_control}.


\section{Results}\label{sec:results}
The performance of the RL agent, MPC, and RL-Guided MPC is evaluated in both the deterministic and stochastic environments.
The results for an additional variant of RL-Guided MPC are also presented in the deterministic environment.
For each environment, the three methods are compared based on the $EPI$, defined in (\ref{eq:EPI}), and the cumulative reward, equal to (\ref{eq:optimization_problem:objective_function}) with the oppposite sign.
For the figures in this section, we use the name `RL-MPC' for the RL-Guided MPC for the sake of brevity.

\subsection{Deterministic case}
\begin{figure}[H]
	\begin{center}
		\includegraphics[width=8.4cm]{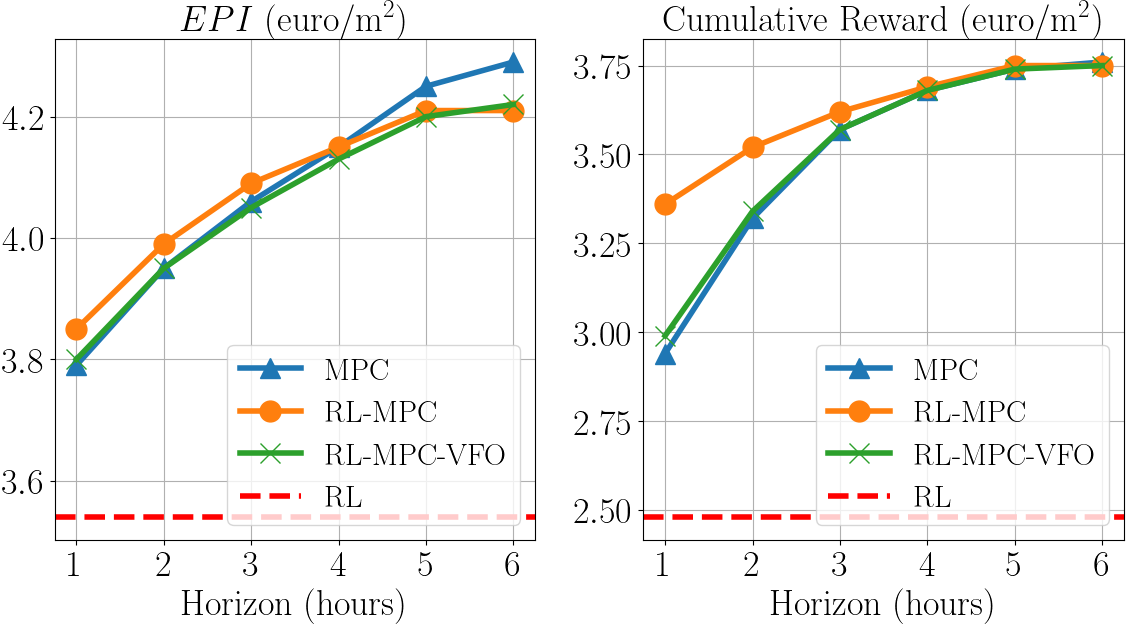}    
		\caption{Comparison of $EPI$ and cumulative reward between RL, MPC, and RL-Guided MPC, at different prediction horizons, for the deterministic case.}\label{fig:deterministic_results}
	\end{center}
\end{figure}

For the deterministic case, an additional variant of RL-Guided MPC is evaluated.
This variant, denoted as `RL-MPC-VFO', is identical to RL-MPC, except that the terminal region constraint is not included in the optimization problem.
The simulation results for the deterministic environment are shown in Figure~\ref{fig:deterministic_results}.
In the $EPI$ comparison, MPC and RL-Guided MPC exhibit similar performance across all prediction horizons, though RL-Guided MPC slightly outperforms MPC at shorter horizons. 
In contrast, the RL agent performs significantly worse than both methods.
The RL-Guided MPC variant without the terminal region constraint performs similarly to MPC.
\\
\begin{figure}[H]
	\begin{center}
		\includegraphics[width=8.4cm]{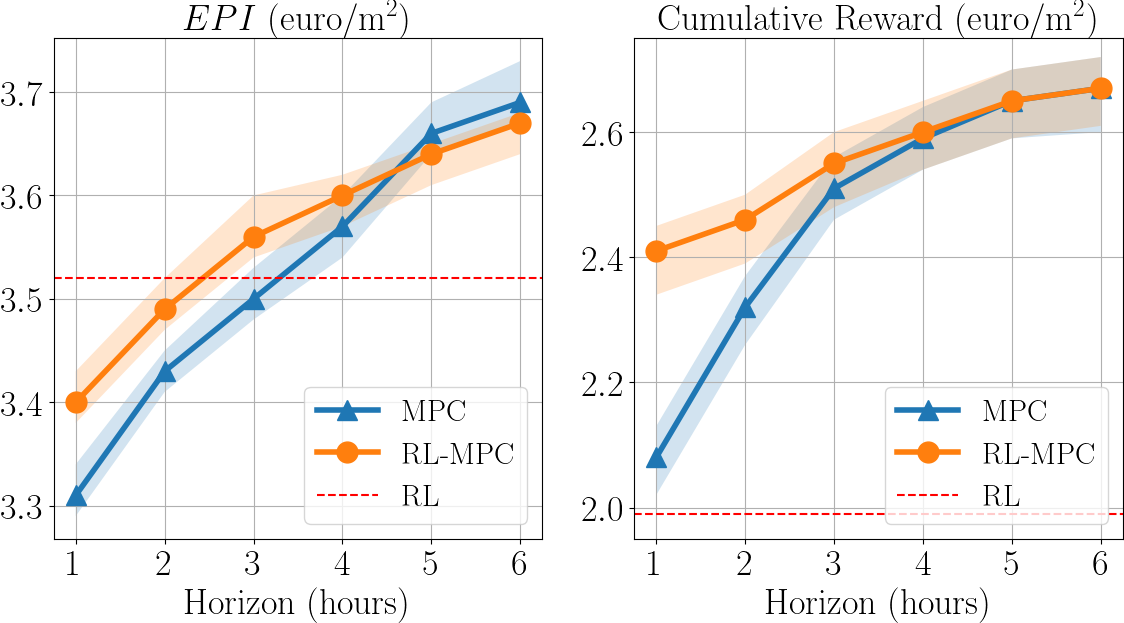}    
		\caption{Comparison of $EPI$ and cumulative reward between RL, MPC, and RL-Guided MPC, at different prediction horizons, for the stochastic case.}\label{fig:stochastic_results}
	\end{center}
\end{figure}
Although comparing controllers using $EPI$ is informative, all controllers aim to maximize the cumulative reward.
In the cumulative reward comparison, a clear distinction emerges between MPC and RL-Guided MPC at lower prediction horizons, where RL-Guided MPC significantly outperforms MPC. 
This advantage stems from the RL-Guided MPC's ability to leverage the learned policy to account for long-term effects. 
However, as the prediction horizon increases, the difference in cumulative rewards diminishes, and beyond four hours, both methods exhibit similar performance. 
Meanwhile, the RL agent continues to perform significantly worse due to its difficulty in handling constraints effectively. 
In contrast, MPC and RL-Guided MPC are explicitly designed to manage constraints, allowing them to operate near constraint boundaries without incurring penalties.
The RL-Guided MPC variant without the terminal region constraint performs similarly to MPC.
This result highlights the importance of the terminal region constraint in the RL-Guided MPC framework, which enables the agent to guide the optimization problem without relying solely on the learned cost function approximator.
\subsection{Stochastic case}
For the stochastic case, 30 different realizations of parameter $\hat p$, in (\ref{eq:uncertain_parameters}), were generated.
For each realization, the RL agent, MPC, and RL-Guided MPC were evaluated and the results were averaged.
The outcomes are shown in Figure~\ref{fig:stochastic_results}, where the shaded regions indicate the range between the minimum and maximum values.
The results for the stochastic case show similar trends to the deterministic case, with some small differences.
Considering the $EPI$ comparison, RL-Guided MPC outperforms MPC at short prediction horizons.
Unlike the deterministic case, the RL agent surpasses both MPC and RL-Guided MPC in terms of $EPI$ at short horizons. 
Consistent with the deterministic case, the cumulative reward analysis confirms RL-Guided MPC's advantage over MPC at shorter 
horizons, particularly showing a notable improvement at a one-hour horizon.

\section{Conclusions}\label{sec:conclusions}
This paper introduced the RL-Guided MPC framework for autonomous greenhouse control, combining the strengths of RL and MPC. 
By training an RL agent and integrating its learned policy into the MPC formulation through the construciton of a terminal cost and terminal region constraint, the proposed approach leverages RL's robustness to uncertainties and MPC's constraint-handling capabilities. 
Simulations in deterministic and stochastic environments demonstrated that RL-Guided MPC outperforms standalone RL and MPC with shorter prediction horizons. 
These results emphasize the value of embedding learned policies into MPC for enhanced performance, without 
the need to extend prediction horizons. 
However, it is important to note that the RL policy trained in this study did not perform as well as expected, 
with the cumulative reward being lower than that of the other methods, for every prediction horizon. 
In future work, we aim to investigate the potential impact of a better-performing policy, which we expect to result in substantial improvements.
Additionally, using a policy derived from RL algorithms is not the only approach.
There are various ways to obtain a suitable policy. 
For greenhouse control, for example, an expert knowledge rule-based controller could serve as an alternative policy, 
potentially offering valuable insights and improved performance.

\section*{Acknowledgements}
The authors thank Sjoerd Boersma and Bart van Laatum for their valuable feedback on this work.
The authors acknowledge the use of computational resources of the DelftBlue supercomputer, 
provided by Delft High Performance Computing Centre (https://www.tudelft.nl/dhpc).

\bibliography{ifacconf}

\end{document}